# Co-benefits of Agricultural Diversification and Technology for Food and Nutrition Security in China

Thomas Cherico Wanger[1,2,3,4,5,*], Estelle Raveloaritiana[2,3,x], Siyan Zeng [2,3,x], Haixiu Gao[6,7,x], Xueqing He[8,4,x], Yiwen Shao[9,10,x], Panlong Wu[11,x], Kris A.G. Wyckhuys[12,13,14,x], Wenwu, Zhou[15,4,x], Yi Zou[8,4,x], Zengrong Zhu[15,16,4,x], Ling Li[1,3], Haiyan Cen[9,10,4,x], Yunhui Liu[17,x], Shenggen Fan[7,*]

**Authors affiliations:**

[1] Research Center for Industries of the Future, Westlake University, Hangzhou, China.

[2] Sustainable Agricultural Systems & Engineering Laboratory, School of Engineering, Westlake University, Hangzhou, China.

[3.] Key Laboratory of Coastal Environment and Resources of Zhejiang Province & School of Engineering, Westlake University, Hangzhou, China.

[4.] China Rice Network, Hangzhou, China.

[5.] Agroecology, University of Göttingen; Göttingen, Germany.

[6.] Institute of Food and Nutrition Development, Ministry of Agriculture and Rural Affairs, Beijing, China

[7.] Academy of Global Food Economics and Policy, China Agricultural University, Beijing, China

[8.] Department of Health and Environmental Sciences, School of Science, Xi'an Jiaotong-Liverpool University, Suzhou, China.

[9.] College of Biosystems Engineering and Food Science, Zhejiang University, Hangzhou, China

[10.] Key Laboratory of Spectroscopy Sensing, Ministry of Agriculture and Rural Affairs, Hangzhou, China.

[11.] School of Ecology and Environment, Inner Mongolia University, China

[12.] Chrysalis Consulting, Danang, Vietnam

[13.] Institute of Plant Protection, China Academy of Agricultural Sciences (CAAS), Beijing, China

[14] School of Biological Sciences, University of Queensland, Saint Lucia, Australia

[15] Ministry of Agricultural and Rural Affairs Key Laboratory of Molecular Biology of Crop Pathogens, Institute of Insect Science, Zhejiang University, Hangzhou 310058, China.

[16] State Key Laboratory of Rice Biology, Institute of Insect Sciences, Zhejiang University, Hangzhou, China

[17] College of Resources and Environmental Sciences, China Agriculture University, China

[*] Main Corresponding Authors: tomcwanger@gmail.com (TCW); s.fan@cau.edu.cn (SF); [x] Corresponding Authors: all other authors

China is the leading producer of maize, wheat, potato, and rice, and has successfully implemented sustainable development programs related to agriculture[1]. Sustainable agriculture has been promoted to achieve national food security targets such as food self sufficiency through the 'well-facilitated farmland construction' or WFFC approach. The WFFC is introduced in China's current national 10-year plan [2] to consolidate traditional and often small-scale farmlands into large and simplified production areas to maximize technology-based automation, and improve soil fertility and productivity. However, research suggests that diversified and smaller farms faciliate ecosystem services, can improve yield resilience, raise production quality, defuse human health threats, and increase farm profitability[3,4]. As the early stage implementation of WFFC has not considered ecological farmland improvements, it may miss long-term environmental benefits for biodiversity conservation and ecosystem service preservation conducive to yields[5]. Moreover, the nutritional status in China has changed significantly in recent decades with undernutrition being dramatically reduced, but the prevalence of overweight, obesity, and chronic diseases has increased. While a strategic choice and management of crop and livestock species can improve nutrition, the environmental and production benefits of agricultural diversification are currently not well interlinked with China's food and nutrition security discussions. In addition, most nationally grown crops rely upon insect-mediated pollination and biological pest control services that are enhanced through agricultural diversification. Lastly, the role of agricultural technology (*sensu* [6]) for socioeconomic benefits and the link with diversified agricultural production may provide vast benefits for food security. Here, we focus on the opportunities and co-benefits of agricultural diversification and technology innovations to advance food and nutrition security in China through ecosystem service and yield benefits (Fig. 1). Our applied five-point research agenda (Tab. 1) can provide evidence-based opportunities to support China in reaching its ambitious food security targets through agricultural diversification with global ramifications.

**China's food security discussion**

China is the world's second most populous country and aims to feed approximately 20% of the human population on 9% of the world's farmland[7]. Over the past 50 years, agricultural productivity in China has grown steadily at 4.6% per year [8]. An increasing population and wealth have also shifted Chinese diets from plant to meat-based foods, which has resulted in sourcing agricultural products from countries in the Americas, Asia and Africa [9]. More than 70% of agricultural fields in China are below 0.64 ha, which is much smaller than the world's average smallholder farm size of 2 ha[10]. The 200-300 million smallholder farmers in China are highly susceptible to environmental impacts[11], suggesting that climate or pest-related supply fluctuations on smallholders and their crop production will have strong implications for local and global food security. Impacts of pest invasions or extended droughts on agricultural production - as witnessed in and beyond China in recent years, have led to commodity price volatility and a perceived need to maintain the use of agrochemical inputs.

Overall, China's sustainable food production is challenged by an overreliance on chemical inputs [12]. The annual nitrogen application in China is on average four times higher than globally (305 kg/ha and 74 kg/ha, respectively), but nitrogen use efficiency (nitrogen use / harvested product) is only 60% in China (0.25 and 0.42, respectively). From the 1950s to the beginning of the 21st century, annual pesticide production in China has increased from 500 tons to 929,000 tons. As pesticide overuse increases the carbon footprint of food, drives environmental pollution, compromises food safety, and causes biodiversity loss, the Chinese government has now made great efforts to mitigate pesticide overuse and promote alternative management schemes. Agricultural diversification in conventional cropping systems can enhance ecosystem services such as biodiversity, pollination, and pest control by 40%, 32%, and 23%, respectively, without yield losses[13]. However, diversification practices without non-chemical inputs can entail yield trade-offs, and, hence, a solid scientific base to balance environmental sustainability and food security is critical.

**Opportunities for agricultural diversification in China**

Agricultural diversification practices include crop rotation, intercropping, crop genetic diversification, non-crop diversification like flower strips, integrated crop-animal systems and agroforestry. These practices can provide a range of water-, biodiversity-, and soil-related ecosystem services[13,14]. For instance in rice production, a synthesis of 40 years of data from China and globally has shown that diversification improves financial profitability, biodiversity and pest control while not affecting yields compared to non-diversified rice paddies[15]. Wheat/maize intercropping systems have 22% higher grain yields than monocultures and greater year-to-year yield stability[16]. In combination with other non-chemical tactics such as releases of biological control agents or the incorporation of organic matter and biofertilizer, diversification in China's cropping systems may benefit yields, yield resilience, and the environment.

Diversification can also help to adjust China's WFFC to fully capitalize on ecosystem service benefits to expand grain production and strengthen national food security as part of the farmland consolidation strategy on 72 million hectares by 2030 [2]. The existing WFFC focuses on improving land productivity by consolidating small fields into larger ones, increasing soil fertility, and improving accessibility and irrigation. Yet, by increasing field size and reducing semi-natural habitats[5], land consolidation will likely reduce landscape heterogeneity along with biodiversity and jeopardize ecosystem services. Through diversification, WFFC can fullfil the requirements on ecological conservation and environmental protection, for instance, by avoiding pesticide application and by conserving or establishing semi-natural habitats at the landscape scale. Integrating diversification practices in China's major policies will leverage the potential of on- and off-farm biodiversity to sustain and enhance food security across the country's diverse geography and ecosystems (Fig. 1) [5].

China has implemented strict policies to ensure sufficient farmland for rising food demand. Diversification strategies do not conflict with these policies. Instead, diversified systems offer long-term win-win outcomes between yields and other ecosystems services compared to simplified systems[17]. In rice production systems, yield and other services such as nutrient cycling and soil fertility can be optimized at the same time in 81% of all known cases, but trade-offs exist for yield and climate regulation[15]. For specific diversification practices, such as organic farming, uptake is limited by strong yield – ecosystem service tradeoffs that are amplified by nitrogen inputs[18]. For specific guidance, spatially explicit approaches can help to better understand where diversification can help to optimize environmental and socioeconomic outcomes in China[19]. Overall, understanding trade-offs in diversification implementation requires more empirical work to then guide targeted implementation (Tab. 1).

## *Pollination and biocontrol to stabilize yields and reduce inputs*

Agricultural diversification can enhance ecosystem services such as pollination and pest control that subsequently increase crop yield, enhance yield resilience, and ultimately underpin food security. In China, insect pollination is of great importance for food security and is particularly critical for diverse and high-quality foods like fruits and vegetables[20]. For example, insect pollination contributes to a respective 57 % and 25 t/ha increase in fruit set and yield in apple production. With the improvement of human living standards in China, demand for pollinator-dependent crops such as fruits and vegetables has recently increased, and their cultivation area has grown by a respective 45% and 15% over the past 20 years (Fig.2). China has also earned global acclaim for its biological pest control approaches. For instance, industrial-scale production of minute parasitic wasps (*Trichogramma* spp.) enables cost-effective pest control on millions of hectares of maize, rice, sugarcane, and fruit crops[21]. Moreover, the application of an insect-killing fungus (*Metarhizium anisopliae*) can effectively control China's most prominent rice pests while safeguarding beneficial organisms and on-farm biota[22]. Thus, pollination and biological pest control are pivotal to food security and nutritional diversity in China with biological pest control creating large potential to reduce chemical inputs.

*Implications of diversification for food and nutrition security*

Agricultural production is key to enhancing nutrition, but the pathways from production management decisions to nutritional benefits are complicated. At the landscape scale, higher diversity can provide more nutrients[23] and more diverse food to subsistence farmers, particularly in rural areas. At the farm level, production diversity is mostly an effective lever to improve the dietary diversity and nutrition of smallholder farmers in developing countries[24]. Furthermore, the link between nutritional status and specific food groups is important. For instance, high livestock diversity is associated with a positive nutritional status among children and adolescents[25]. Moreover, crop diversity highly relates to food accessibility and dietary diversity[26]. In China, studies confirm the positive link between agricultural production diversity, dietary diversity in rural residents and to be conducive for an increase of protein intake[27].

While the links between diverse production and improved nutrition become better understood globally, more research is needed to connect production and nutritional status in China (Fig. 1). Especially, because the Chinese government has placed great emphasis on food self-sufficiency and agricultural development in rural areas, as highlighted in China's most important agricultural policy (i.e., No.1 central document; 1CD) [28]. Specific requirements were then introduced in important agriculture-related meetings and documents to satisfy people's increasingly diversified food needs through diversified food supply. For example, the Central Rural Work Conference of China in 2015 and 2017 highlighted "the concept of establishing big agriculture and big food". In 2023, 1CD emphasized again "establishing a big food concept and accelerating the construction of a diversified food supply system"[29]. The methods of diversifying agricultural production, such as crop-livestock systems and agroforestry, have been clearly mentioned. It is important to note that the big food concept is not explicitly linked to the WFFC yet. Thus, agricultural diversification has become a prime option to ensure China's food and nutrition security.

**Co-benefits of agricultural technology and diversification for food and nutrition security**

In achieving sufficient food production, China has substantially increased its agricultural mechanization through strategies such as WFFC. While to date already one-third of agricultural production has attained a high level of mechanization, the majority still depends on traditional operations by smallholder farmers.

Adoption of agricultural mechanization faces two major challenges in China linked to agricultural diversification. First, application of agricultural machinery is widely anticipated to be incompatible with highly fragmented landscapes, large spatial and temporal differences, and small production scale[30]. This is, because mechanization of crop production is increasingly perceived as leading to intensive farming in simplified and centralized systems. Such production systems cannot provide sufficiently diverse products of better quality, and higher nutritional value. Second, even after WFFC consolidated small-scale farms into large production areas, there is an insufficient supply of and a lack of training on agricultural machinery in both consolidated and fragmented production areas. According to Chinese-type modernization, an aging farming population in China must reduce labour intensity while enhancing production efficiency and boosting unit earnings to avoid farmland abandonment [31]. With China's sustainability strategy in mind, these expectations cannot be met merely in simplified and mechanized production systems, but must be supported by advances in digital agriculture in diversified landscapes.

Digital or smart agriculture refers to data-driven management of agricultural production systems, where characteristics of crops, soil, and climate are collected by different technologies to support farmers' decision-making and along the agricultural value chain. At the pre-production stage, optical sensors and multiscale phenotyping platforms can help quantifying plant traits for genetic selection[32], both of new or local crop varieties. During the production stage, digital technologies are used to quantify plant photosynthetic function, dynamic growth, and nutrient cycling based on the biochemical characteristics of leaves[33]. Advances in embedded systems allows standardized pollinator and pestcontrol monitoring for direct yield improvements[34]. For the post-production period, digital technologies can help with pre-sale quality control in both, the conventional and organic agricultural market[35]. There is, hence, not necessarily a discrepancy but rather a synergy between diversified and sustainable farming and the use of new technology but the links between them are not well established.

The Chinese government has been passing policies that integrate digital technologies in sustainable agricultural production (2024 1CD) [36]. Digital agriculture to support or replace mechanization through automated farm management is still in its infancy, but an increase of up to 40% by 2025 is expected [37]. This increase of digitization and integration with diversification practices requires careful consideration of China's diverse geography and ecosystems. For instance, the agriculture Internet of Things (aIOT) uses sensor data, drones, and specialized sensors to integrate real-time data on crop phenology, pest incidences, and weather patterns, suitable for large-scale rice production with different crop varieties. Agricultural robotics can advance automated seeding, non-chemical weeding, and harvesting in any large scale crop production such as apple orchards planted as agroforestry systems. Online market places faciliate market access for farmers in rural areas and provide the convenience of selling premium-priced and traceable products that can be bought online[38]. To faciliate Chinese-type modernization, the government could further support the use of these digital agriculture approaches in diversified farming landscapes[5] by providing financial incentives for technology development and training to faciliate implementation[39]. Overall, mechanization approaches may interfere with agricultural production diversification, but new technology innovations in diverse and small farms likey will amplify the benefits for ecosystem service preservation, yield stability, and climate resilience[40].

**Outlook**

During the past two decades, China has drawn a red-line policy to protect the quality of its 120 million hectares of arable land and many important developments are already emerging to improve ecological resilience in the agricultural sector[11]. In the major rice-producing areas in Northeast China, conservation tillage will be applied in about 70% of total cultivated land by returning straw to fields in 2025 [41]. Moreover, synthetic pesticides will increasingly make room for biological pest control alternatives in food and horticultural croplands. Smart agriculture and extended use of drip irrigation are being promoted to raise resource use efficiency across the nation's farming areas. The establishment of 780 new science and technology centers will help transform China's smallholder agriculture from an intense to a sustainable intensification model[39]. Equally, to reach ambitious goals in pesticide phase-down, several specialized 'ecological pest management' research units have been established over the past years[42]. Major benefits will also be reaped by tactically integrating the above practices with natural, bred, or engineered varieties resistant to pests and extreme climate conditions. In general, recent policies have advocated for a multidisciplinary research environment, which now requires proper incentive structures to reward scientists for collaboration instead of competition. In addition, to better link agricultural production diversity, nutrition status, and technology innovation for Chinese food security, we propose a five-point research agenda that requires strong interdisciplinary and multi-stakeholder collaborations (Tab. 1). Overall, a cornucopia of human and environmental health benefits awaits to be enjoyed, and effective policies in China are key to ensure that science translates into practice.

**Author Contributions**

Conceptualization, coordination, funding acquisition: TCW; Methodology: TCW;

**Figures**

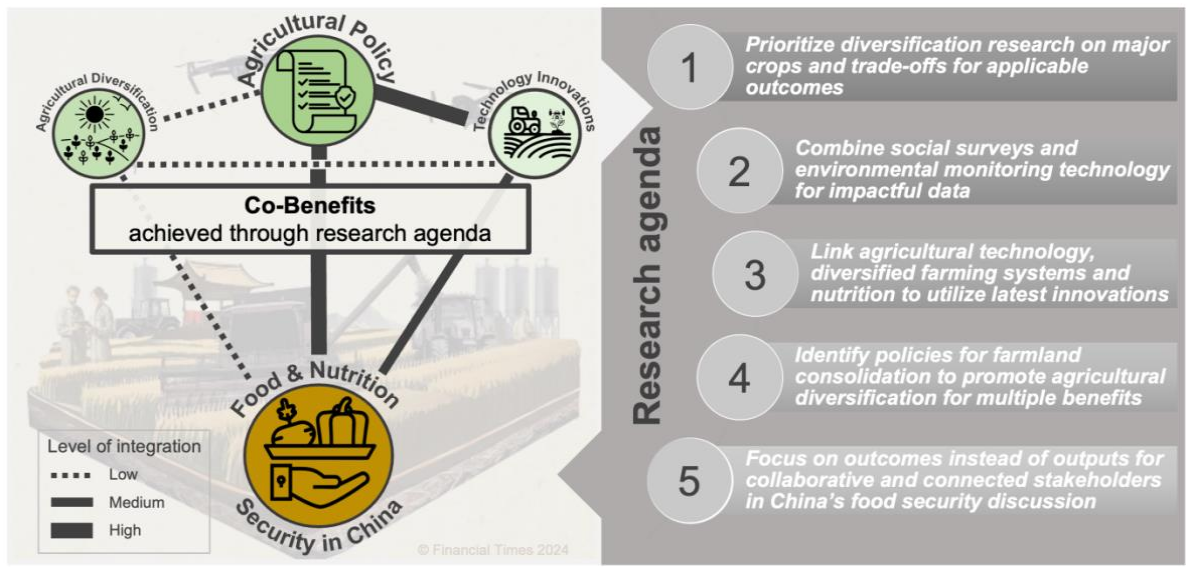

**Fig. 1.** Co-benefits of agricultural diversification and technology integration for food and nutrition security in China. The "Level of Integration" between the four components 'Agricultural Policy', 'Agricultural Diversification', 'Technology Innovations' & 'Food and Nutrition Security in China' is based on published evidence and expert opinions from the author team.

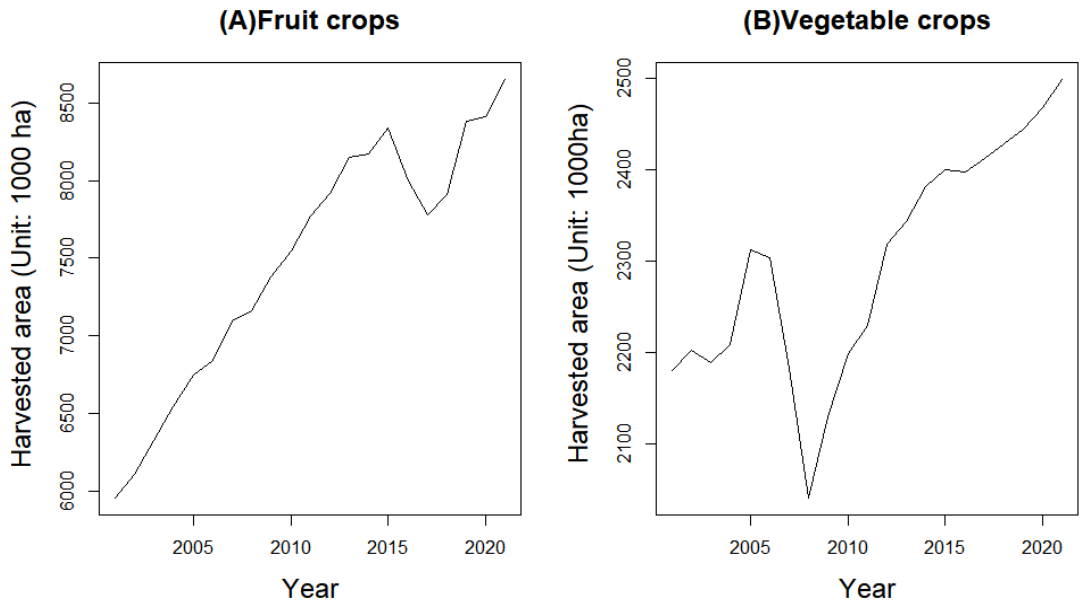

**Fig. 2.** Progressive increase in the harvested area of (A) fruit and (B) vegetable crops that rely upon insect pollination in China from 2001 to 2021; data source: https://www.fao.org/statistics/data-collection.

**Tables**

**Table 1.** A practical research agenda for integrating agricultural diversification and technology innovations to achieving Chinese food nutrition security goals.

| Research Agenda Item | Details | References |
|---|---|---|
| ***1. Prioritize diversification research on major crops and trade-offs between yield, financial profitability, ecosystem services, and nutrition for applicable outcomes in China's multifunctional agricultural landscapes.*** | i. Location-specific case studies on staple food and cash crops such as rice, wheat, cotton, rubber and cocoa for implementation-based recommendations;<br>ii. Links to sufficient high-quality and safe food are critical;<br>iii. Urban and rural gradients at the farm and landscape scale must be reflected. | *(T1,T2,T3-6)* |
| ***2. Combine social science survey techniques and environmental monitoring technology to effectively link different stakeholder perspectives with empirical environmental and nutritional data.*** | i. Aim should be to understand farmers' awareness of and option on i) diversified systems; ii) their role in providing diverse diets; and iii) adoption of technology in diversified systems;<br>ii. Resulting datasets should be accessible and long-term (ideally >5 years). | *(T2,T5,T7,T8)* |
| ***3. Link agricultural technology in diversified farming systems with nutrition security to fully utilize the best innovations across disciplines for sustainable food production in China and globally.*** | i. Agricultural automation with advanced sensing and artificial intelligence amongst many technologies are promising and not necessarily in stark contrast to sustainable agriculture;<br>ii. Mainstreaming this aspect is critical to overcome the perceived hurdles of implementing combined new technology measures in sustainable and diversified agricultural systems.<br>iii. Dedicated research and development is required on specialized agricultural machinery (e.g., for mountainous regions and in the marine environment), considering sustainable development. | *(T9-12)* |
| ***4. Identify and develop policy measures and strategies especially for farmland consolidation to promote agricultural diversification for meeting environmental and nutritional objectives.*** | i. Above-ground (e.g., crop rotation and intercropping of grains and legumes including crucifer cover crops) and below-ground diversification (e.g., inoculation, conservation tillage) in compatibility with well-facilitated farmland construction should be targeted;<br>ii. policies should enable technology use in combination with applied diversification practices as local governments adapt policies to regional conditions while ensuring detailed and context-specific implementation;<br>iii. environmental, nutritional, and other socioeconomic objectives should build on dialogue between policymakers, farmers, | *(T1,T3,T10,T13-15)* |
| | researchers, and technology providers to ensure that diversification strategies are practical and beneficial. | |
| **5. Focus on outcomes with impact instead of outputs by metrics for truly collaborative research and connected stakeholders in the food security discussion in China.** | i. The right incentive structures are needed to transition away from output (i.e., counting first and corresponding authorships of publications) towards outcome focused (i.e., how to further advance sustainable food production in China based on interdisciplinary insights from publications) collaborative research for impact;<br>ii. Facilitate establishment of multidisciplinary research and innovation centers that foster a culture of multistakeholder collaboration;<br>iii. We make this publication 'count' for all authors by making everyone a corresponding author. | *(T16)* |
| **References** | T1. G. Tamburini, *et al.*, *Sci Adv* **6**, eaba1715 (2020).<br>T2. T. C. Wanger, *et al.*, *Nat. Ecol. Evol.* **4**, 1150–1152 (2020).<br>T3. T. Tscharntke, *et al., Trends Ecol Evol* **36**, 919–930 (2021).<br>T4. X. He, *et al., Nat food* **4**, 788–796 (2023).<br>T5. G. Gurr, *et al., Nat Plants* **2**, 16014 (2016).<br>T6. D. Kleijn, *et al., Nat Commun* **6**, 7414 (2015).<br>T7. M. H. Ruckelshaus, *et al., Trends Ecol Evol* **35**, 407–414(2020).<br>T8. D. Kleijn, *et al., Trends Ecol Evol* **34**, 154–166 (2019).<br>T9. S. Zeng, *et al.*, *Commun Earth Environ* **4**, 183 (2023). | |